\documentclass[12pt]{article}
\usepackage[dvips]{graphics}
\usepackage{amssymb}
\usepackage[utf8]{inputenc}
\usepackage{xcolor}
\newcommand{\be}{ \begin{equation} }
\newcommand{\ee}{\end{equation}} 
\begin{document} 
\def\theequation{\arabic{section}.\arabic{equation}} 
\begin{titlepage} 
\title{Perfect fluid solutions of Brans-Dicke and 
$f(R)$ cosmology}

\author{Dilek K. \c{C}iftci$^{a,b}$\footnote{email addresses: 
dkazici@ubishops.ca, dkazici@nku.edu.tr}  and Valerio 
Faraoni$^{b}$\footnote{email address: vfaraoni@ubishops.ca}\\ \\  
{\small   $^a$ Department of Physics, Nam{\i}k Kemal University, 
Tekirda\u {g}, Turkey}\\ 
{\small  $^b$Department of Physics \& Astronomy and {\it STAR} Research 
Cluster}\\
{\small  Bishop's University, 2600 College Street,  
Sherbrooke, Qu\'{e}bec, Canada J1M~1Z7}  
}

\date{} 
\maketitle 
\vspace*{1truecm} 
\begin{abstract} 

Brans-Dicke cosmology with an (inverse) power-law potential is revisited 
in the light of modern quintessence and inflation models. A simple 
ansatz relating scale factor and scalar field recovers most of the known 
solutions and generates new ones. A phase space interpretation of 
the ansatz is provided and these universes are mapped into 
solutions of $f(R)$ cosmology.

\end{abstract} 
\vspace*{1truecm} 
\end{titlepage}

\def\theequation{\arabic{section}.\arabic{equation}}


\section{Introduction}
\label{sec:1}
\setcounter{equation}{0}

There are currently many theoretical and experimental investigations of 
possible deviations of gravity from Einstein's theory in cosmology, 
black holes, gravitational waves, and the dynamics of galaxies and galaxy 
clusters \cite{Padilla}. The standard $\Lambda$CDM model of cosmology 
requires the introduction of a completely {\em ad hoc} dark energy 
accounting for 
70\% of the energy content of the universe \cite{AmendolaTsujikawa}. An 
alternative to dark energy 
consists of modifing gravity at large scales, which has led to 
contemplating many theories of gravity, and especially the so-called 
$f(R)$ class \cite{reviews1, reviews2, reviews3, CapozzielloFaraoni}. 
These are essentially 
scalar-tensor 
theories, and scalar-tensor gravity is the prototypical alternative to 
General Relativity (GR) which introduces only an extra scalar degree of 
freedom. The simplest scalar-tensor gravity was proposed by 
Brans and Dicke in 1961 \cite{BD} and was later generalized 
\cite{ST1, ST2, ST3} and is 
still the subject of active research. In any case, although Solar System 
tests do not show deviations from GR, gravity is tested poorly in many 
regimes while it is not tested at all in 
others \cite{Berti2015, Psaltis}, and there is plenty of room for 
deviations from GR. 
Apart from the motivation arising from cosmology, attempts to unify 
gravity and quantum mechanics invariably produce deviations from GR in the 
form of extra degrees of freedom, higher order field equations, and extra 
tensors in the Einstein-Hilbert action, so it is expected that eventually 
GR fails at some energy scale. Indeed, the simplest 
 bosonic string theory reduces to a Brans-Dicke
theory with coupling parameter $\omega=-1$ in the low-energy limit 
\cite{string1, string2}. 
Motivated by the flourishing of cosmological models in alternative 
gravity and especially in $f(R)$ and other scalar-tensor gravities, we 
revisit the simplest incarnation, Brans-Dicke cosmology with a scalar 
field potential. There are now over five decades of research on this 
subject but not many analytical solutions are known which describe  
spatially homogeneous and 
isotropic cosmology (see \cite{mybook, FujiiMaeda} for partial reviews). 
Contrary to the original Brans-Dicke theory, in 
which the extra gravitational scalar field was 
free and 
massless, we study the situation in which it acquires a power-law or 
inverse power-law potential, which has now been included in a large 
number of cosmological scenarios related to inflation or to the present 
acceleration of the universe \cite{extended1, extended2, extended3,  
extended4, hyperextended1, hyperextended2, hyperextended3, 
hyperextended4, hyperextended5, hyperextended6, STquintessence1, 
STquintessence2, STquintessence3, STquintessence4, 
STquintessence5, STquintessence6, STquintessence7, 
STquintessence8, STquintessence9, STquintessence10, 
STquintessence11, STquintessence12, STquintessence13, 
STquintessence14, STquintessence15, STquintessence16, 
STquintessence17, STquintessence18, STquintessence19, 
STquintessence20, STquintessence21,  
STquintessence23, STquintessence24, STquintessence25, 
STquintessence26, STquintessence27, STquintessence28, 
STquintessence29, STquintessence30, STquintessence31, 
STquintessence32, STquintessence33, mybook, FujiiMaeda, 
CapozzielloFaraoni}. As shown in the following 
sections, most 
of 
the known solutions of spatially homogeneous and isotropic Brans-Dicke   
cosmology can be derived from a simple ansatz, which allows one to uncover new solutions of this theory with exponential scale factor and scalar 
field. A geometric interpretation of this ansatz in terms of the 
geometry of the 
phase space of the solutions is proposed in Sec.~\ref{sec:3}. The old and 
new solutions of Brans-Dicke cosmology with potential 
are then mapped into solutions of $f(R)$ cosmology in Sec.~\ref{sec:4}.

We begin with the Brans-Dicke action\footnote{We follow the notation of 
Ref.~\cite{Wald}.} \cite{BD}
\be
S_{BD}=\int d^4x \sqrt{-g} \left[ \phi R -\frac{\omega}{\phi} \, g^{ab} 
\nabla_a \phi \nabla_b \phi -V(\phi) \right] +S_{(m)} \,,\label{1}
\ee
where $R$ is the Ricci scalar, $\phi$ is the Brans-Dicke scalar field, 
$\omega$ is the constant Brans-Dicke coupling, $V(\phi)$ is a 
potential for the Brans-Dicke field, and $S_{(m)}$ is the matter action. 
Here we assume a power-law 
or inverse power-law potential  
\be
V(\phi)  =V_0 \phi^{\beta} \,, \label{potential}
\ee 
with $V_0$ and $\beta$ constants and 
$V_0\geq 0$. This form of the potential is motivated by large bodies of 
literature on inflation \cite{extended1,  extended2, 
extended3, extended4, hyperextended1, hyperextended2, hyperextended3,
hyperextended4, hyperextended5, hyperextended6} and 
quintessence 
\cite{STquintessence1, STquintessence2, STquintessence3, STquintessence4,
STquintessence5, STquintessence6, STquintessence7,
STquintessence8, STquintessence9, STquintessence10,
STquintessence11, STquintessence12, STquintessence13,
STquintessence14, STquintessence15, STquintessence16,
STquintessence17, STquintessence18, STquintessence19,
STquintessence20, STquintessence21, 
STquintessence23, STquintessence24, STquintessence25,
STquintessence26, STquintessence27, STquintessence28}. 

The Brans-Dicke field equations in the Jordan frame are
\begin{eqnarray}
R_{ab}-\frac{1}{2}\, g_{ab}R &=& \frac{8\pi}{\phi} \, T_{ab} 
+\frac{\omega}{\phi^2} \left( \nabla_a \phi \nabla_b \phi
-\frac{1}{2} \, g_{ab} \nabla^c \phi \nabla_c \phi \right) 
\nonumber\\
&&\nonumber\\
&\, & +\frac{1}{\phi}  \left( \nabla_a \nabla_b \phi
- g_{ab} \Box \phi \right)  -\frac{V}{2\phi} \, g_{ab} \,, \label{BD1} \\
&&\nonumber\\
\Box \phi &=& \frac{1}{2\omega+3} \left( 8\pi T +\phi \, \frac{dV}{d\phi} 
-2V \right) \,,\label{BD2}
\end{eqnarray}
where $\nabla_a$ is the covariant derivative operator, $\Box \equiv 
g^{ab}\nabla_a\nabla_b$, $T_{ab}$ is the 
energy-momentum tensor of ordinary matter, and 
$T \equiv {T^a}_a$ is its trace. In the following we assume that 
$\omega \neq -3/2$ and that matter consists of  a 
perfect fluid with stress-energy tensor
\be 
T_{ab}=\left(P+\rho \right) u_a u_b +Pg_{ab} \label{perfectfluid}
\ee
(where $u^a$ is the fluid 4-velocity) and with barotropic, linear, and 
constant equation of state 
\be
P=\left( \gamma -1 \right) \rho \,, \;\;\;\;\;\; \gamma=\mbox{const.}
\label{eos}
\ee
relating the energy density $\rho$ and pressure $P$. 
A cosmological constant can be introduced in the theory by 
considering a linear potential $V=\Lambda \phi $. In fact, since the 
vacuum action of GR contains the combination $R-\Lambda$, 
and the Brans-Dicke field $\phi$ multiplies $R$ in the Brans-Dicke action, 
the natural way of introducing a cosmological constant in Brans-Dicke 
theory is through the combination $\phi \left(R-\Lambda \right)$, which is 
equivalent to introducing a linear potential $V=\Lambda \phi$. Or, 
considering the Brans-Dicke field equation~(\ref{BD1}), it is obvious that  
adding a term $\Lambda g_{ab}$ to the left hand side is equivalent to 
inserting a  linear potential $V=\Lambda \phi$ in the right hand side.

The parameters of the theory are $\left( \omega, \beta, V_0, 
\gamma \right)$. We now specialize to spatially homogeneous and isotropic 
Brans-Dicke cosmology, with the geometry given by the 
Friedmann-Lema\^{i}tre-Robertson-Walker (FLRW) line element
\be
ds^2=-dt^2 +a^2(t) \left( \frac{dr^2}{1-kr^2} +r^2 d\Omega_{(2)}^2 
\right) \label{FLRW}
\ee
in comoving coordinates, where  $k$ is the curvature index and $ 
d\Omega_{(2)}^2=d\theta^2 +\sin^2 \theta \, d\varphi^2$ is the line 
element on the unit 2-sphere. The equations of Brans-Dicke cosmology with 
the perfect fluid~(\ref{perfectfluid}) and (\ref{eos}) consist of the 
Friedmann, acceleration, and scalar field  equations
\begin{eqnarray}
 H^2 &=& \frac{8\pi \rho}{3\phi} +\frac{\omega}{6} \, 
\frac{\dot{\phi}^2}{\phi^2} -H\, \frac{\dot{\phi}}{\phi} -\frac{k}{a^2} 
+\frac{V}{6\phi} \,, \label{5}\\
&&\nonumber\\
\dot{H} & = & \frac{-8\pi}{\left(2\omega+3\right)\phi} \left[ \left( 
\omega+2 \right)\rho +\omega P \right] -\frac{\omega}{2} \, 
\frac{\dot{\phi}^2}{\phi^2} +2H \, \frac{\dot{\phi}}{\phi} + \frac{k}{a^2} 
\nonumber\\
&&\nonumber\\
& \, & + \frac{1}{ 2 \left(2\omega+3\right)\phi} \left( \phi\, 
\frac{dV}{d\phi} - 
2V \right) \,,\label{6}\\
&&\nonumber\\
\ddot{\phi} &+& 3H \dot{\phi} = \frac{1}{ 2\omega+3} \left[ 8\pi 
\left(\rho-3P \right)-\phi \, \frac{dV}{d\phi} +2V \right] \,,\label{7}
\end{eqnarray} 
respectively, where $H \equiv \dot{a}/a$ is the Hubble parameter and an 
overdot denotes 
differentiation with respect to the comoving time $t$. In addition, the 
covariant conservation equation $\nabla^b T_{ab}=0$ yields
\be
\dot{\rho}+3H\left(P+\rho \right)=0 
\ee
which, using the equation of state~(\ref{eos}), is immediately integrated 
to
\be
\rho(a) = \frac{\rho_0}{a^{3\gamma}} \,, \label{integrated}
\ee
where $\rho_0$ is a non-negative integration constant.

\section{New and old solutions}
\label{sec:2}

Since the early days of scalar-tensor gravity, various authors have looked 
for FLRW solutions of this class of theories in the power-law form, $a(t) 
\propto t^q$ and $\phi(t) \propto t^s$.  Here we search for solutions 
satisfying the ansatz $\phi(t)=\phi_0 \, a^p$. {Scaling and power-law 
relations are ubiquitous in physics \cite{Pietroneroetal2001}, biology 
\cite{HuxleyTeissier1936, Gayon2000}, geophysics and glaciology 
\cite{Horton1945, Strahler1957, DoddsRothman2000, BahrPfefferKaser2015}, 
and various natural sciences, and it is rather natural to investigate such 
relations in cosmology. The fairly large literature studying 
cosmological 
power-law solutions $a(t) \sim t^q, \phi(t) \sim t^r$ is mostly well 
motivated from the physical point of view. The ansatz $\phi =\phi_0 a^p$ 
reproduces almost all these power-law solutions. The 
physical meaning of this ansatz resides in the fact 
that the effective gravitational coupling strength becomes $G_{eff} \sim 
\phi^{-1} \sim a^{-p}$ and the ansatz relates directly the strength of 
gravity with the cosmic scale factor.  If $p>0$, the 
effective gravitational coupling decreases as the universe expands, while 
$G_{eff}$ increases if $p<0$. These two behaviours are separated by GR, 
which corresponds to $p=0$. The assumption $\phi(t) = \phi_0 a^p$ can be 
rewritten in a covariant way as $u^c \nabla_c \phi/\phi = - p \, 
\Theta/3$, 
where $\Theta$ is the expansion of the congruence of observers comoving 
with the cosmic fluid, which have timelike 4-tangent $u^c$. According 
to the comoving time associated with these observers, $ 
\dot{G}_{eff}/G_{eff}=- \dot{\phi}/\phi$. The ansatz $\phi =\phi_0 a^p$  
offers a self-consistent scenario realizing the  assumption 
$\dot{G}/G \sim H$ used in analyses of the variation of the gravitational 
coupling. This assumption is often made on a purely phenomenological 
basis and corresponds to the (rather vague) idea that $G$ varies on a  
cosmological time scale, in order to place experimental or 
observational constraints on the variation of $G$ (cf., {\em e.g.},  
Refs.~\cite{WillLivingReviews, AccettaKraussRomanelli}). If the evolution 
of $ G_{eff} $ and 
that of the scale factor are not 
tied together directly, as in our ansatz in the context of scalar-tensor 
gravity, it is difficult to see how the desired phenomenological relation  
$\dot{G}/G \sim H$ can be obtained, and how it can be obtained in a  
covariant way.

In our study we first want  
to recover the known power-law solutions, hence we begin by assuming that   
\begin{eqnarray}
a(t) &= & a_0 t^q \,,\label{11} \\
&&\nonumber\\
\phi(t) &=& \phi_0 a^p \,, \label{12}
\end{eqnarray}
where $q(\omega, \beta, \gamma) $ and $p(\omega, \beta, \gamma)$ are 
exponents to be determined as functions of the parameters of the theory 
and $ a_0, \phi_0 $, and $\rho_0$ in Eq.~(\ref{integrated}) are constants. 
We require that $p \neq 0$ because otherwise one has a constant scalar 
field which reduces Brans-Dicke theory to GR, which is a
trivial situation in our context.

The form~(\ref{11}) and (\ref{12}) of the solutions of Brans-Dicke 
cosmology to which we restrict is sufficiently general to allow us to 
recover a host of classic solutions. 
Later on, we will relax the assumption~(\ref{11}) but we will keep the 
ansatz~(\ref{12}) finding new exponential, instead of power-law, solutions 
for spatially flat FLRW universes. An 
interpretation of the ansatz~(\ref{12}) in the phase space of 
the solutions is given in Sec.~\ref{sec:3}.

 There are two possible approaches to the solution of 
Eqs.~(\ref{5})-(\ref{7}).  In the first approach, the first step of the 
solution process consists of finding 
exponents 
$q(\omega, \beta, \gamma) $ and $p(\omega, \beta, \gamma)$ that solve all 
three field  equations~(\ref{5})-(\ref{7}). This is obtained by matching 
the powers 
of the comoving time $t$ in these equations. The Friedmann 
equation~(\ref{5}) gives 
\be
\frac{q^2}{t^2} \left( 1+p-\frac{\omega p^2}{6}  \right) +\frac{k}{a_0^2\ 
t^{2q} } 
=\frac{8\pi \rho_0}{3a_0^{p+3\gamma} \phi_0} \, 
\frac{1}{t^{q(p+3\gamma)}}+\frac{V_0\phi_0^{\beta-1}}{6\ a_0^{p(1-\beta)}} \, 
\frac{1}{t^{pq(1-\beta)}} \, .
\ee
Matching the powers of $t$ in each term yields the following relations:
\begin{eqnarray}
\mbox{if} \ \  k\neq 0 \ , \quad & & \mbox{it must be}  \quad
q=1 \,;\label{15}\\
&&\nonumber\\
\mbox{if} \ \  \rho_0\neq 0 \ , \quad & & \mbox{it must be} \quad 
q\left( p+3\gamma \right)=2 \, ; \label{16}\\
&&\nonumber\\
\mbox{if} \ \  V_0\neq 0 \ , \quad & & \mbox{it must be} \quad
pq\left( 1-\beta \right)=2 \ .\label{17}
\end{eqnarray}
The second possible approach to solving Eqs.~(\ref{5})-(\ref{7}) 
consists of noting 
that some of the four terms in the Friedmann equation~(\ref{5}) could 
balance each other, without having to match all the powers of $t$. 
However, the acceleration and field equations impose further constraints 
and, in practice, this method does not lead to new solutions with respect 
to those obtained with the first method (the details of this second 
approach are presented in \ref{appendix}). Let us continue, 
therefore, with the first solution method. The second step of this 
process consists of 
taking the functions $ q(\omega, \beta, \gamma)$ and $ p(\omega, \beta, 
\gamma)$ found in the previous step (if they exist) and of 
determining the various integration constants $a_0, \phi_0,  \rho_0$ as 
functions of  $\left( \omega, \beta, \gamma, V_0 , p, q\right)$. 
Using again the Friedmann equation~(\ref{5}), computer 
algebra provides the value of the integration constant for the 
density\footnote{We are not aware of general formulae in the literature 
analogous to ~(\ref{eq:2.7})-(\ref{eq:2.9}), which provide the values 
of these integration constants in terms of the theory parameters 
$\omega \,, \gamma \,, V_0 \,, \beta$ and of the exponents $q$ and $p$. 
This is possibly related to the fact that computer algebra 
was not available at the time of early explorations of scalar-tensor 
gravity and of its solutions.}
$\rho_0$ as a function of the 
other two integration constants $\phi_0$ and 
$a_0$ as
\be
 \rho_0= \frac{\phi_0 a_0^{p+3\gamma}}{16\pi} \left[ \frac{6k}{a_0^2} -V_0 
 \phi_0^{\beta-1} a_0^{p(\beta-1)} +q^2 \left( 6+6p-\omega p^2\right) 
 \right] \,. \label{eq:2.7}
\ee
Then the scalar field equation~(\ref{7}) provides  the 
integration constant $\phi_0$ appearing in the Brans-Dicke field as
\begin{eqnarray}
&&\phi_0 = \frac{1}{a_0^p \left[ (2\beta-3\gamma)V_0 \right]^{ 
\frac{1}{\beta-1}} } \nonumber\\
&&\nonumber\\
& & \cdot \left[ 
\frac{6k(4-3\gamma)}{a_0^2} +2pq\left(1-pq -3q\right) 
\left( 2\omega+3\right) +q^2\left( 6+6p-\omega p^2 \right) \left(4-3\gamma 
\right)  \right]^{\frac{1}{\beta-1}} \nonumber\\
&& \label{eq:2.8}
\end{eqnarray}
for $V_0 \neq 0$.
The acceleration equation~(\ref{6}) provides another such relation for the 
integration constant $a_0$:
\begin{eqnarray}
&&a_0^2 = \frac{2k}{q} \cdot \nonumber\\
&&\nonumber\\
&& \cdot \left[ 
\frac{3\gamma(\beta-1)-2\beta}{2\gamma(p\omega-3) 
+2\beta(p+2)+q\gamma(1-\beta)(6+6p-p^2\omega) 
+2q(p\beta+3\gamma)(1-p-p\omega)}\right] \,. \nonumber\\
&& \label{eq:2.9}
\end{eqnarray}
If instead $V_0=0$, there is no such constraint on 
$\phi_0$ and the expression 
$$
\frac{6k(4-3\gamma)}{a_0^2} +2pq\left(1-pq -3q\right) 
\left( 2\omega+3\right) +q^2\left( 6+6p-\omega p^2 \right) \left(4-3\gamma 
\right) 
$$ 
must vanish.

In spite of the simplicity introduced by the assumptions~(\ref{11}) 
and~(\ref{12}), the field equations (\ref{5})-(\ref{7}) are still 
non-linear and quite 
involved and it is 
convenient to analyze the various possibilities separately.

\subsection{$k=0 \,, V_0=0 \,, \rho_0=0$}
\label{subsec:2.1}

In this vacuum none of the constraints~(\ref{15})-(\ref{17}) between 
$p, q, \beta$, and $\gamma$ apply. The 
Friedmann equation~(\ref{5}) becomes simply
\be
 6+6p- \omega p^2 =0 
\ee
and, in a non-static universe, it provides the values of $p$
\be
p_{\pm} = \frac{ 3 \pm \sqrt{3(2\omega +3)}}{\omega } \,. \label{19}
\ee
The acceleration equation~(\ref{6}) then gives
\be
q_{\pm}=\frac{\omega}{3(\omega+1)\pm\sqrt{3(2\omega+3)}}\label{19q} \,.
\ee
Using the values~(\ref{19}), (\ref{19q}) and Eq.~(\ref{12}), one concludes that
\begin{eqnarray}
a(t) &=&a_0 \, t^{\frac{2}{p(p\omega-4)} }= a_0\, t^{\frac{ \omega}{3(\omega+1) \pm \sqrt{3(2\omega+3)} } } 
\,,\label{23}\\ 
&&\nonumber\\
\phi(t) &=& \phi_*\, t^{\frac{1 \pm \sqrt{3(2\omega+3)}}{3\omega+4} }\, ,  
\label{24}
\end{eqnarray}
where $\phi_*=\phi_0 a_0^{\frac{3\pm\sqrt{3(2\omega+3)}}{\omega}}\, .$ This is the classic O'Hanlon and Tupper vacuum solution of 
Brans-Dicke 
cosmology with free scalar field and  
$\omega>-3/2 $, $\omega \neq -4/3, 0$, describing a spatially flat FLRW 
universe \cite{OHanlonTupper}. In this case $p$ and $q$ depend only 
on the Brans-Dicke coupling $\omega$, while the constants $a_0$ and    
$\phi_0 $  are not constrained.


\subsection{$k=0 \,, V_0=0 \,, \rho_0 \neq 0$}
\label{subsec:2.2}

In this non-vacuum case, only the constraint~(\ref{16}) between $p$ and 
$q$ must 
hold, and 
this equation is all the information that can be obtained by matching 
powers of $t$ in the field equations. Equation~(\ref{12}) then yields 
\be
a(t)=a_0 t^{ \frac{2}{p+3\gamma}} \,, \;\;\;\;\;\;\;
\phi(t)=\phi_0\,a_0^p\, t^{ \frac{2p}{p+3\gamma}}\, \equiv\,  
\phi_*t^{\frac{2p}{p+3\gamma}}  \,,  \label{questa}
\ee
but the possible values of $q$ and $p$ are still unknown. In order to 
determine them, one substitutes Eq.~(\ref{questa}) in the Friedmann 
equation~(\ref{5}), obtaining
\be
\rho_0 = \frac{ \left( 6+6p-\omega p^2\right)}{4\pi \left( p+3\gamma 
\right)^2}\, \phi_0 a_0^{3\gamma+p} \,,\label{eq:2.20}
\ee
and substituting this in the acceleration equation (\ref{6}), one obtains 
an algebraic equation for $p$  with roots
\be
p_{+}= \frac{3\gamma-4}{\omega(\gamma-2)-1} \,, \;\;\;\;\;\;\;\;
p_{-}= \frac{3}{\omega} \,.
\ee
The other field equation~(\ref{7}) must also be satisfied, and it is 
satisfied by the root $p_{+}$ but not by $p_{-}$. Therefore, using 
Eq.~(\ref{16}), one concludes that the only solution of the desired form 
corresponding to a spatially flat FLRW universe with 
free Brans-Dicke scalar and with perfect fluid is
\begin{eqnarray} 
a(t) &=& a_0 t^{\frac{ 2\left[ \omega (\gamma-2)-1\right]}{ 3\omega\gamma 
(\gamma-2) -4}}  
\,,\label{29}\\ 
&&\nonumber\\
\phi(t) &=& \phi_* t^{\frac{2(3\gamma-4)}{ 3\omega\gamma 
(\gamma-2)-4}}  \,, \label{30}\\
&&\nonumber\\
\rho(t) &=& \rho_*  t^{-\frac{ 6\gamma \left[ \omega 
(\gamma-2)-1\right]}{3\omega\gamma(\gamma-2)-4}} \,
\end{eqnarray}
for $3\omega\gamma (\gamma-2) \neq 4$ and with 
\be
\phi_* = \phi_0 a_0^{ \frac{3\gamma-4}{\omega(\gamma -2)-1} } \,, 
\;\;\;\;\;\; \rho_*=\frac{\rho_0}{a_0^{3\gamma}} \,\label{30-b}. 
\ee 
This is recognized as the Nariai solution  
\cite{Nariai1, Nariai2}.
The power $q$ of the scale factor $a(t) \simeq t^q$ is independent of the 
Brans-Dicke coupling $\omega$ if $\gamma=2$ or if $\gamma=4/3$.  The 
constants 
$a_0$ and $\phi_0$ are not constrained and $\rho_0$ is given in terms of 
them and of $\omega, \gamma,p$ by Eq.~(\ref{eq:2.20}).

\subsubsection{Exponential solutions}

For $k=0, V_0=0, \rho_0\neq 0$, there are 
expanding/contracting de Sitter 
spaces with exponential scalar fields. Assuming $H=$~const., the Friedmann 
equation~(\ref{5}) becomes
\be
\left( 6+6p-\omega p^2 \right) H^2 =\frac{16\pi \rho_0}{\phi_0 
a^{p+3\gamma}}  \,
\ee
and it can be satisfied for $\rho_0\neq 0$ and constant $H$  only if 
\be
p = -3\gamma \,,
\ee
which yields
\be
H^2 = \frac{ 16\pi \rho_0}{3\phi_0 \left[ 2-3\gamma (2+\omega 
\gamma ) \right] } \,.
\ee
In order to satisfy the Friedmann and acceleration equations, it must 
also be 
\be
\gamma=1\pm \sqrt{ \frac{3\omega+4}{3\omega} } \;\;\;\;\;\;\; 
\mbox{or}\;\;\;\;\;\;\; \gamma=-\, \frac{1}{\omega} \,.
\ee
The second value of $\gamma$, however, does not satisfy the scalar field 
equation and is discarded. The remaining value of $\gamma$ gives 
the solution
\begin{eqnarray}
a(t)&=& a_0\exp({H\ t})  \nonumber\\ 
& = & a_0 \exp\left\{ \pm \left[ 
\frac{8\pi \rho_0}{3\phi_0 \left[ 
-(3\omega+4) \mp 3(\omega+1) \sqrt{ \frac{3\omega+4}{3\omega} } \, \right] 
} \right]^{1/2} \, t \right\} \,, \\
&&\nonumber\\
\phi(t) &=& \phi_* \exp{(pHt)} \nonumber\\
&&\nonumber\\ 
& = & \phi_* \exp\left\{ \mp \left( 
1\pm \sqrt{ \frac{3\omega+4}{3\omega}} \right)\left[  \frac{24\pi \rho_0  }{\phi_0 \left[ 
-(3\omega+4)\mp 3(\omega+1)\sqrt{ \frac{3\omega+4}{3\omega}} \right]} 
\right]^{1/2} \, 
t \right\}\nonumber \,,\\
&&\\
\rho(t)&=& \rho_*\exp{(-3\gamma H t)} \nonumber\\ 
&&\nonumber\\ 
& = & \rho_* \exp\left\{  \mp \left( 1 \pm \sqrt{ 
\frac{3\omega+4}{3\omega}} \right) 
\left[  
\frac{24\pi \rho_0}{ \phi_0 \left[
-(3\omega+4) \mp 3(\omega+1)\sqrt{ \frac{3\omega+4}{3\omega} } \,\right] }  
\right]^{1/2} \, t \right\}  \,,\nonumber\\
&&
\end{eqnarray}
where $\phi_* =\phi_0 a_0^{ -3\left( 1\pm \sqrt{ \frac{3\omega+4}{3\omega}} 
\right) }$ and $\rho_* = \rho_0 a_0^{ -3\left( 1\pm \sqrt{ 
\frac{3\omega+4}{3\omega}} \right) }$.
Ordinary matter has 
$\gamma \geq 1$, which corresponds to $p=-3\gamma <0 $ and $G_{eff} \sim 
a^{-p}$ increases on a cosmological time scale as the universe expands.


\subsection{$k=0 \,, V_0\neq 0 \,, \rho_0 = 0$}
\label{subsec:2.3}

Of the three constraints, only~(\ref{17}) must be satisfied in this 
vacuum. One finds 
\be
 p=\frac{2(2-\beta)}{1+2\omega+\beta} \label{this-p}
\ee
and, therefore,
\begin{eqnarray}
a(t) &=& a_0 t^{ \frac{1+2\omega+\beta}{(2-\beta)(1-\beta)} }\,,\\
&&\nonumber\\
\phi(t) &=& \phi_* t^{ \frac{2}{1-\beta} } \,,
\end{eqnarray}
while $a_0$ is not constrained, $\phi_*=\phi_0\ a_0^{\ p} $ and  
\be
\phi_0=a_0^{\frac{2(\beta-2)}{1+2\omega+\beta}}  \left[ \frac{2(2\omega+3)\left[ 6\omega -(\beta+1)(\beta-5) 
	\right] 
}{V_0 (\beta-1)^2 (\beta-2)^2} \right]^{ \frac{1}{\beta-1} } \,.
\ee

\subsubsection{Exponential solutions}

For $k=0$, $\rho_0=0$, and $V_0 \neq 0$, there are 
exponential solutions which are expanding 
or contracting de Sitter spaces with $H=$~const. and 
$\dot{\phi}/\phi=pH=$~const.  Assuming $H=$~const., the Friedmann 
equation~(\ref{5}) reduces to
\be
\left( 6+6p-\omega p^2 \right) H^2 =\frac{V_0}{\phi_0^{1-\beta}  
a^{p(1-\beta)} }   \,,
\ee
which can be satisfied for $V_0\neq 0$ and constant $H$ only if $
\beta=1 $. Further, the acceleration equation~(\ref{6}) implies that
\be 
p=\frac{1}{\omega +1} \,.
\ee
The solutions of the desired form, therefore, exist only if 
$V(\phi)=V_0\phi$ (which corresponds
to a cosmological constant) and are  
\begin{eqnarray}
a(t) &=& a_0\exp{(Ht)} \nonumber\\
&&\nonumber\\
&=& a_0 \exp \left\{ \pm \left(\omega+1 \right)\left[ \frac{V_0}{ 
(2\omega+3)(3\omega+4)} \right]^{1/2} \, t \right\}  \,,\\
&&\nonumber\\
\phi (t)&=& \phi_* \exp{(pHt)} \nonumber\\
&&\nonumber\\
&=& \phi_* \exp 
\left\{\pm  \left[ \frac{V_0}{ 
(2\omega+3)(3\omega+4)} \right]^{1/2} \, t \right\} \,,
\end{eqnarray}
where $\phi_*=\phi_0 a_0^{\ \frac{1}{\omega +1} }$.
Contrary to GR, the scalar field of these de Sitter spaces is not 
constant. These solutions are well known attractors in the phase space of 
Brans-Dicke cosmology, with an attraction basin which is wide but does 
not span all of the phase space \cite{81, 744, 544}. For $\omega=-1$ 
there are no simultaneous solutions of Eqs.~(\ref{5})-(\ref{7}).


\subsection{$k=0 \,, V_0\neq 0 \,, \rho_0 \neq 0$}
\label{subsec:2.4}

In this non-vacuum case, the two constraints (\ref{16}) and (\ref{17}) 
must be satisfied simultaneously. There are no solutions if $\beta=1$ 
while, if $\beta\neq 1$ there is the unique solution 
$\left(q, p \right) = \left( \frac{2\beta}{3\gamma(\beta-1)}, 
\frac{-3\gamma}{\beta} \right)$ for $\gamma\neq 0$. Therefore, the solution 
is
\begin{eqnarray}
a(t) &=& a_0 t^{\frac{2\beta}{3\gamma(\beta-1)}} \,,\\
&&\nonumber\\
\phi(t) &=& \phi_* t^{ \frac{2}{1-\beta}} \,,\\
&&\nonumber\\
\rho(t) &=& \rho_* t^{\frac{2\beta}{1-\beta}} \,.
\end{eqnarray}
The integration constants $a_0$, $\phi_0$, and $\rho_0$ are related by
\begin{eqnarray}
\phi_0 &=&a_0^{\frac{3\gamma}{\beta}} \left[  \frac{4 }{3V_0} 
\frac{ 
	\left( 1+2\omega-\omega\gamma\right)3\gamma-\left( 
	3\gamma-4\right)\beta }{ 
	\gamma^2(\beta-1)^2} \right]^{ \frac{1}{\beta-1}}\nonumber\, , \\&&\nonumber\\\rho_0 &=& \frac{\phi_0 a_0^{\frac{3\gamma(\beta-1)}{\beta} }  }{12\pi} 
\frac{ \left( 2\beta^2-6\beta\gamma -3\omega \gamma^2 \right)}{\gamma^2 
\left( \beta-1\right)^2} 
-\frac{V_0\phi_0^{\beta} }{16\pi} \,,
\end{eqnarray}
where $\phi_*=\phi_0\, a_0^p $ and $\rho_*=\rho_0\, a_0^{-3\gamma}$. 

In the special case $\gamma=0$ excluded thus far and corresponding to the 
equation of state $P=-\rho=\mbox{ constant}$, the 
constraint on $q$ becomes $q= 2/p$  and $ \beta=0$. The field 
equations are satisfied by $p=\frac{4}{2\omega+1}$ and the solution 
becomes 
\begin{eqnarray}
a(t) &=& a_0 t^{\omega+\frac{1}{2}} \,,\\
&&\nonumber\\
\phi(t) &=& \phi_* t^2 \,,\\
&&\nonumber\\
\rho (t)&=&-P(t)
=-\frac{V_0}{16\pi}+\frac{\phi_0 a_0^{\frac{4}{2\omega+1}} 
(2\omega+3)(6\omega+5)}{32\pi}\, ,
\end{eqnarray}
where $\phi_*=\phi_0\, a_0^{\frac{4}{2\omega+1}}$, and $\omega \neq -1/2$.

\subsubsection{Exponential solutions} 

The situation $k=0 \,, V_0\neq 0 \,, \rho_0 \neq 0$ admits exponential 
solutions. Assuming $H=$~const., the Friedmann equation~(\ref{5}), which  
becomes
\be
\left( 6+6p-\omega p^2 \right) H^2 =\frac{16\pi \rho_0}{\phi_0 
a^{p+3\gamma}} +\frac{V_0}{\phi_0^{1-\beta} a^{p(1-\beta)} } \,,
\ee
can be satisfied only if 
\be
\beta=1 \;\;\;\;\;\;\; \mbox{and} \;\;\;\;\; p=-3\gamma \,,
\ee
which yields the values of the constant Hubble parameter 
\be
H^2 = \frac{ 16\pi \rho_0+V_0\phi_0}{3\phi_0\left[ 2-6\gamma -3\omega 
	\gamma^2)\right] } 
\ee
and the integration constant
\be
\rho_0 = \frac{V_0\phi_0 \left[ 1+3\gamma(\omega+1) \right] }{8\pi 
\left[ -4+3\omega \gamma(\gamma-2)\right] } \,.
\ee
The solutions, therefore, are the expanding or contracting de Sitter 
spaces with exponential Brans-Dicke field
\begin{eqnarray}
a(t)&=& a_0\exp{(\pm Ht)}\nonumber\\
&&\nonumber\\ 
&=& a_0 \exp\left\{ \pm 
\left[ \frac{V_0}{3\left[ 
4-3\omega\gamma\left( \gamma-2\right) \right] }\right]^{1/2} \, t \right\} 
\,,\\
&&\nonumber\\
\phi(t)&=& \phi_*\exp{(\pm pHt)} \nonumber\\  
&&\nonumber\\ 
&=& \phi_* \exp\left\{ \mp 3\gamma \left[ \frac{V_0}{3\left[ 
4-3\omega\gamma\left( \gamma-2\right) \right] } \right]^{1/2} \, t 
\right\}  \,,\\
&&\nonumber\\ 
\rho(t)&=& \rho_*\exp{(\mp 3\gamma Ht)} \nonumber\\ 
&&\nonumber\\ 
&=& \rho_* \exp\left\{ \mp 
3\gamma \left[ \frac{V_0}{3\left[ 
4-3 \omega\gamma\left( \gamma-2\right) \right]} \right]^{1/2} \, t 
\right\}  \,,
\end{eqnarray}
where $ \phi_*=\phi_0\ a_0^p $ and $ \rho_*=\rho_0\, 
a_0^{-3\gamma} $.

We have $\dot{G}_{eff}/G_{eff} =-pH=-\dot{\rho}/\rho$. 
For ordinary matter with $\gamma \geq 1$ it is $p <0$ and gravity 
becomes stronger as the universe expands and the matter fluid dilutes. 


\subsection{$k \neq 0 \,, V_0=0 \,, \rho_0 \neq 0$}
\label{subsec:2.5}

In this non-vacuum case, it is necessarily $ q= 1$ and $p=2-3\gamma$. The 
FLRW solution is 
\begin{eqnarray}
a(t) &=& a_0 t \,,\\
&&\nonumber\\
\phi (t) &=& \phi_* t^{2-3\gamma} \,,\\
&&\nonumber\\
\rho(t) &=& \rho_* t^{-3\gamma} \,,
\end{eqnarray} 
with $\phi_*=\phi_0\, a_0^{2-3\gamma}$,  $\rho_*=\rho_0\, a_0^{-3\gamma}$. 
Here $\phi_0$ is arbitrary (but positive) and the other integration 
constants 
are 
\begin{eqnarray}
a_0 &=& \left[ \frac{2k}{ \omega (\gamma-2)(3\gamma-2)-2} \right]^{1/2} 
\,,\\
&&\nonumber\\
\rho_0 &=& -\, \frac{k\phi_0}{4\pi} \, \frac{ 
(2\omega+3)(3\gamma-2)}{\omega(\gamma-2)(3\gamma-2)-2} \,.
\end{eqnarray} 
The parameters $k, \omega, \gamma$, of course, must lie in a range such 
that $a_0>0$ and $\rho_0>0$.

If $k=-1$ this universe is just Minkowski space in a foliation with 
time-dependent 3-metric and the line element~(\ref{FLRW}) can be reduced 
to the Minkowski one by an appropriate coordinate transformation 
(see, {\em e.g.}, \cite{Slava}). It is not  a trivial Minkowski space 
because 
the effective stress-energy tensor of the free Brans-Dicke scalar cancels 
out the fluid stress-energy tensor in the field equations~(\ref{BD1}) to 
produce flat spacetime. If $k=1$, spacetime is a genuine positively 
curved FLRW manifold.


\subsection{$k \neq 0 \,, V_0 \neq 0 \,, \rho_0 = 0$}
\label{subsec:2.6}

In this vacuum case, the constraints~(\ref{15}) and (\ref{17}) must be 
satisfied simultaneously, producing
\begin{eqnarray}
a(t) &=& a_0 t \,,\\
&&\nonumber\\
\phi(t) &=& \phi_* t^{\frac{2}{1-\beta}} \,,
\end{eqnarray}
where $\phi_*=\phi_0 a_0^{\frac{2}{1-\beta}}$, while the integration constants are
\begin{eqnarray}
a_0 &=& \left[ \frac{ k(\beta-1)^2}{-1+2\omega+\beta(4-\beta)} 
\right]^{1/2} \,,\\
&&\nonumber\\
\phi_0 &=&
\left[ \frac{ 
4a_0^2 (2\omega+3) }{V_0 \left( \beta-1 \right)^2}   \right]^{ 
\frac{1}{\beta-1}}\, \nonumber\\
&&\nonumber\\&=& \left\{ \frac{4k(2\omega+3) }{ V_0 \left[ 
	-1+2\omega+\beta(4-\beta)\right]} \right\}^{ 
\frac{1}{\beta-1}}  \,.
\end{eqnarray}


\subsection{$k \neq 0 \,, V_0 \neq 0 \,, \rho_0 \neq 0$}
\label{subsec:2.7}

In this non-vacuum case, all of the three constraints 
(\ref{15})-(\ref{17}) 
between  $p,q,\beta$, and $\gamma$ must be satisfied 
simultaneously. There are no solutions of the desired form if $\beta=1$. 
If $\beta \neq 1$, it is necessarily $a(t)=a_0 t$, while  
Eq.~(\ref{17}) gives
\be
p=\frac{2}{1-\beta} \,, 
\ee
which does not depend on the Brans-Dicke coupling $\omega$, while 
Eq.~(\ref{16}) yields $p=2-3\gamma$. 
By comparing these two values of $p$ it follows that, once the scalar 
field potential $V(\phi)=V_0 \phi^{\beta}$ is fixed, the perfect fluid 
equation of state is also 
necessarily fixed to
\be
\gamma= \frac{2\beta}{ 3(\beta-1)} \,.
\ee
The solution is the FLRW universe~(\ref{FLRW}) with  scale factor,  
Brans-Dicke field, and fluid energy density   
\begin{eqnarray}
a(t) &=& a_0 t \,,\\
&&\nonumber\\
\phi(t) &=& \phi_* t^{\frac{2}{1-\beta}} \,,\\
&&\nonumber\\
\rho(t) &=& \rho_* t^{\frac{2\beta}{1-\beta}}  
\end{eqnarray}
with $\phi_*=\phi_0\, a_0^{\frac{2}{1-\beta}} $ and $\rho_*=\rho_0\, a_0^{\frac{2\beta}{1-\beta}}$. The integration constants 
are 
\begin{eqnarray}
\phi_0 &=& \left\{ \frac{2a_0^2}{\beta V_0}\left[3+\frac{3k}{a_0^2} +\frac{
2\omega (2\beta-3)}{(\beta-1)^2}\right] \right\}^{ \frac{1}{\beta-1} } 
\,,\\
&&\nonumber\\
\rho_0 &=& -\frac{\phi_0\, a_0^2}{4\pi} \bigg(\frac{2\omega+3}{\beta-1} 
\bigg)+  \frac{V_0\phi_0^\beta}{16\pi}(\beta-1) 
 \,.
\end{eqnarray}


\section{A phase space interpretation}
\label{sec:3}

We now provide a geometric interpretation of the ansatz $\phi(t)=\phi_0 
a^p(t)$ in the phase space of the 
solutions. For simplicity, we restrict to  the simplest situation, 
which corresponds to the parameter 
values $k=0$ and $\rho_0=0$, in which case the dimensionality of the phase 
space reduces to three, thus allowing for an intuitive graphical 
interpretation (a generic description of the phase space of Brans-Dicke 
cosmology was given in Ref.~\cite{AnnPhys}). 

When $k=0$ and in the absence of matter, the scale factor $a(t)$ enters 
the cosmological equations~(\ref{5})-(\ref{7}) only through the 
combination $H 
=\dot{a}/a$ and one can choose as variables the 
Hubble parameter and the Brans-Dicke scalar $\left( H(t), \phi(t) 
\right)$. Then the phase space reduces to $\left( H, \phi , 
\dot{\phi} 
\right)$. The Friedmann equation~(\ref{5}), which is of first order, then 
acts as a constraint  
which forces the orbits of the solutions  to lie on the analogue of the 
``energy surface'' with equation~(\ref{5}), effectively reducing the phase 
space accessible to 
these orbits to a 2-dimensional subset\footnote{We refer to this ``energy 
surface'' in quotation marks because it can be self-intersecting, as in 
the example below, and it is not an embedded hypersurface in the usual 
sense of geometry.} of the 3-dimensional space $\left( 
H, \phi, \dot{\phi} \right)$. Let us examine this ``energy surface'' in 
the simple case $k=0, \rho_0=0$. The constraint equation~(\ref{5}) 
becomes 
\be
H^2=  \frac{\omega}{6} \, \frac{\dot{\phi}^2}{\phi^2} - 
\frac{H\dot{\phi}}{\phi} +\frac{V_0}{6} \, \phi^{\beta-1} \,.
\ee
We can regard it as an algebraic equation for $\dot{\phi}$ and express 
$\dot{\phi}$ as a function of the other two variables $H$ and $\phi$,
\be
\dot{\phi} \left(H, \phi \right)= \left\{ 
\begin{array}{lll} 
\frac{ 3H\phi \pm \sqrt{ 3(2\omega+3)H^2\phi^2 
-\omega V_0\phi^{\beta-1}} }{\omega}  \;\;\;\;\; 
& \mbox{if} &  \omega\neq 0 \,,\\
&&\\  -H\phi + \frac{V_0\phi^{\beta} }{6H}  &  \mbox{if} & \omega = 0 
\,.
\end{array} \right. \label{mecco}
\ee
In general, for $\omega\neq 0$, one or more regions of the 
$\left(H, \phi \right)$ 
plane correspond to a negative argument 
\be
\Delta \equiv 3(2\omega+3)H^2\phi^2 -\omega V_0\phi^{\beta -1}
\ee
of the square root in Eq.~(\ref{mecco}). In this case, the corresponding 
regions in the ``energy surface'' $\left( H, \phi, \dot{\phi}(H, \phi) 
\right)$ are ``holes'' which are avoided by the orbits of the solutions 
(in the sense that no real solutions of the dynamical 
system~(\ref{5})-(\ref{7}) exist 
whose orbits enter these regions). These holes can be infinite or  
semi-infinite. Further, the ``energy surface'' is 
composed of two sheets, corresponding to the positive or negative signs in 
Eq.~(\ref{mecco}), which will be denoted ``upper sheet'' and ``lower 
sheet'' (in keeping with the terminology of Ref.~\cite{AnnPhys}). These 
two sheets join on the boundaries of the ``holes'', which are identified 
by the equation $\Delta=0$.

Looking for solutions satisfying the ansatz $\phi=\phi_0a^p$ means  
intersecting the ``energy surface'' (\ref{mecco}) with the surface of 
equation 
\be
\dot{\phi}\left(H, \phi \right)=pH\phi \,,,
\ee
expressing the assumption that the effective gravitational 
coupling varies as $\dot{G}_{eff}/G_{eff} =-pH$.  
However, the constant $p$ is not assigned {\em a priori}. The problem 
consists of finding {\em simultaneously} values of $p$ for which these 
intersections exist and the intersection curves themselves, which are the 
orbits of the solutions satisfying the desired ansatz.

\begin{figure}[h]
\begin{center}
-\scalebox{0.5}{\rotatebox{-90}{\includegraphics{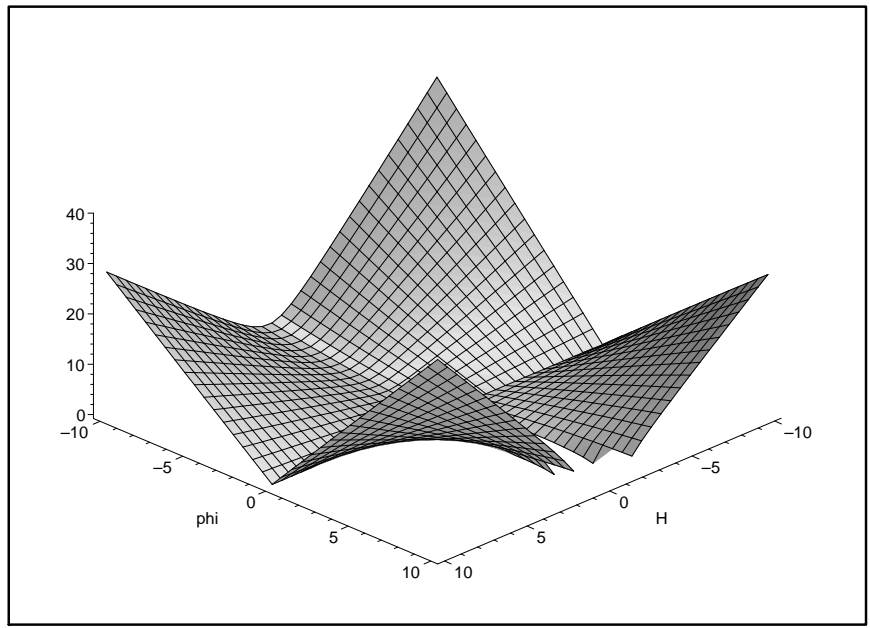}}}
\end{center}
\caption{The upper sheet of the ``energy surface'' corresponding to the 
positive sign in Eq.~(\ref{mecco}).}
\label{upper-sheet}
\end{figure}
As an example, consider the parameter values $\omega=55$ and $\beta=4$ and 
use units in which $V_0$ is unity. Then the region forbidden to the orbits 
of the solutions is given by 
\be
|H|< \sqrt{\frac{55}{339}}\, \phi^{1/2} \,.
\ee 
The two sheets composing the ``energy surface'' have equations
\be
\dot{\phi}_{\pm}\left( H, \phi\right) = \frac{3H\phi}{55} \pm 
\frac{  \sqrt{ 339H^2\phi^2-55\phi^3} }{55} \,.\label{example:sheets}
\ee
Upper sheet, lower sheet, and the ``energy surface'' are 
plotted in Figs.~\ref{upper-sheet}-\ref{both-sheets}. 
\begin{figure}[h]
\begin{center}
\scalebox{0.5}{\rotatebox{-90}{\includegraphics{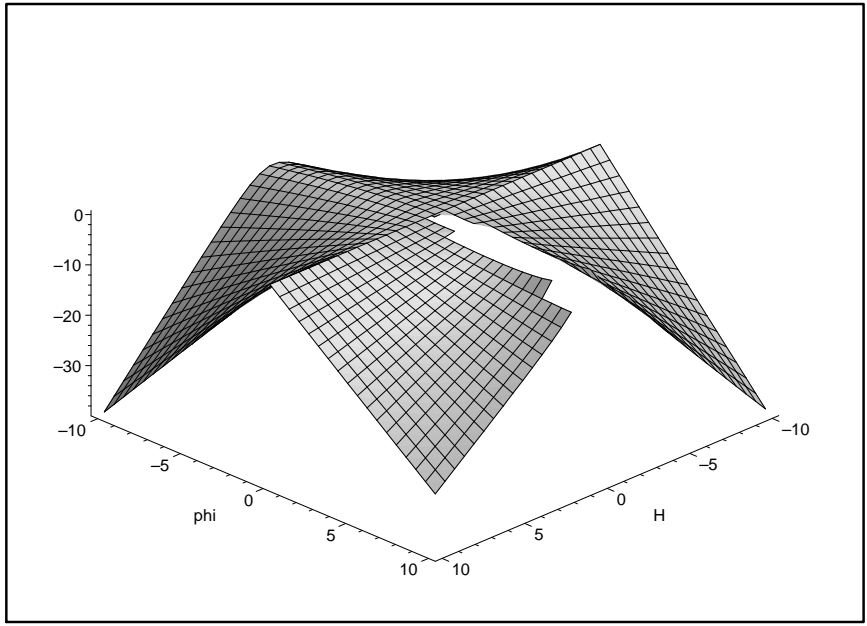}}}
\end{center}
\caption{The lower sheet of the ``energy surface'' corresponding to the
negative sign in Eq.~(\ref{mecco}).}
\label{lower-sheet}
\end{figure}

\begin{figure}[h]
\begin{center}
\scalebox{0.5}{\rotatebox{-90}{\includegraphics{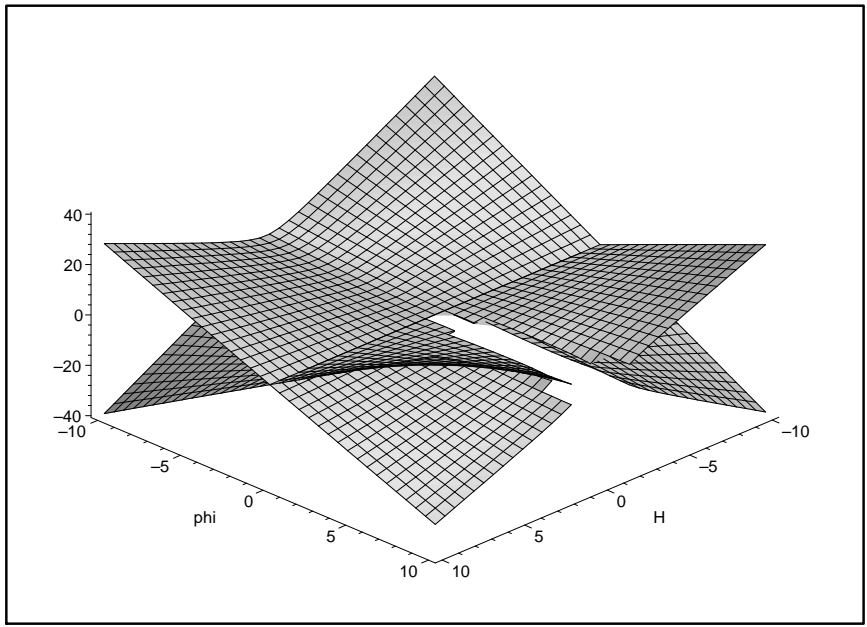}}}
\end{center}
\caption{The ``energy surface''.}
\label{both-sheets}
\end{figure}

The boundary of the hole, where the two sheets join, is the curve whose 
points have coordinates
\be
\left( H, \phi, \dot{\phi} \right) = \left( \pm \sqrt{ \frac{55}{339}} \, 
\phi^{1/2} , \phi,  \pm \sqrt{ \frac{3}{6215}} \phi^{3/2} \right) \,.
\ee 
The intersections between the energy surface~(\ref{example:sheets}) and 
the surface $\dot{\phi}\left( H, \phi \right)=pH\phi $ change as $p$ 
changes. However, only the value of $p$ given by Eq.~(\ref{this-p}), that 
is $ p=-4/115$ for this example, corresponds to actual orbits of the 
solutions 
of 
the dynamical system~(\ref{5})-(\ref{7}).  This ``ansatz surface'' and its 
intersection with 
the ``energy surface'' are plotted in Fig.~\ref{ansatz-surface} and 
Fig.~\ref{intersection}, respectively.
\begin{figure}[h]
\begin{center}
\scalebox{0.5}{\rotatebox{-90}{\includegraphics{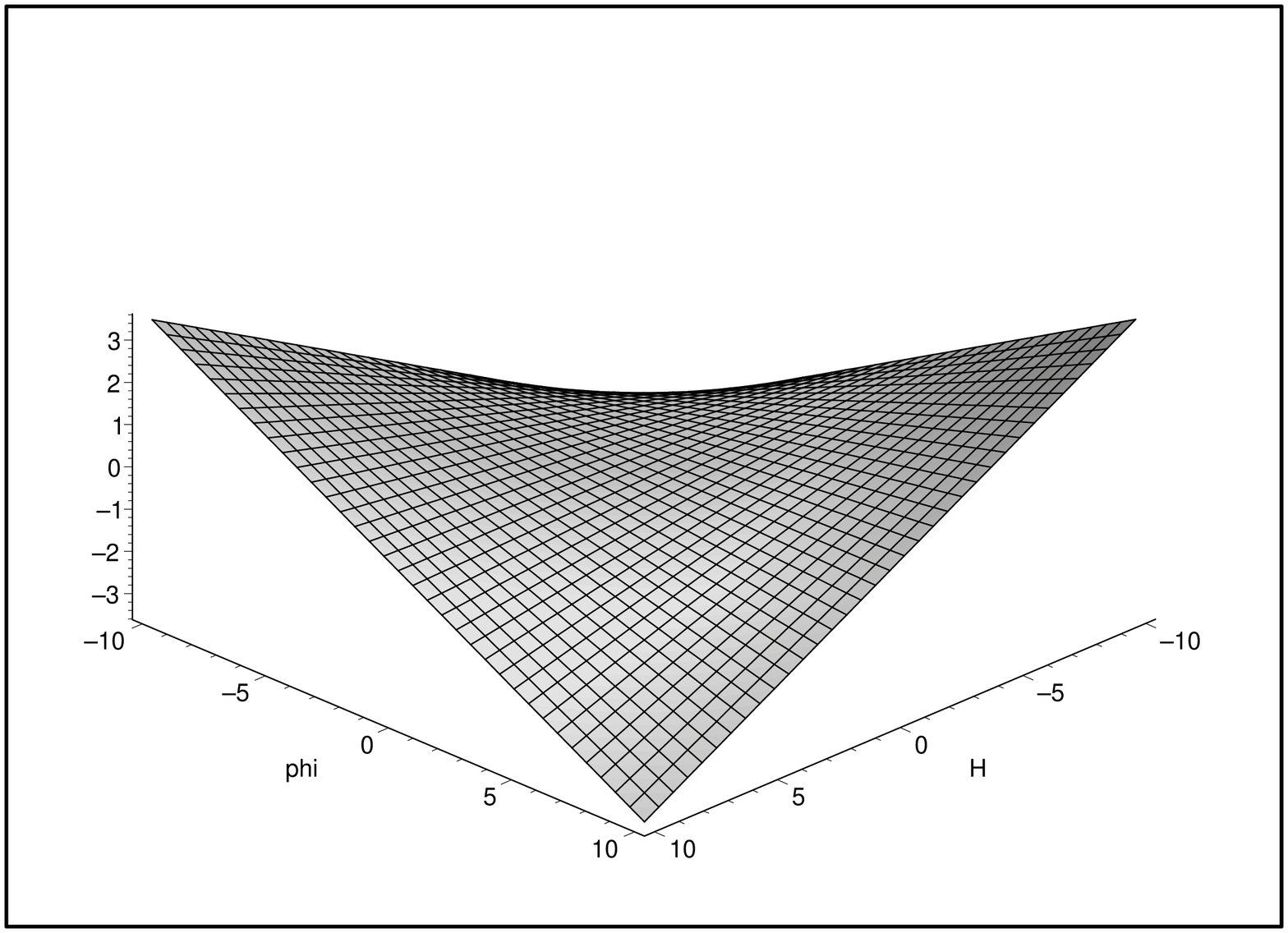}}}
\end{center}
\caption{The ``ansatz surface'' $\dot{\phi}=pH\phi$ corresponding 
to the assumption $\phi=\phi_0 a^p$.}
\label{ansatz-surface}
\end{figure}

\begin{figure}[h]
\begin{center}
\scalebox{0.5}{\rotatebox{-90}{\includegraphics{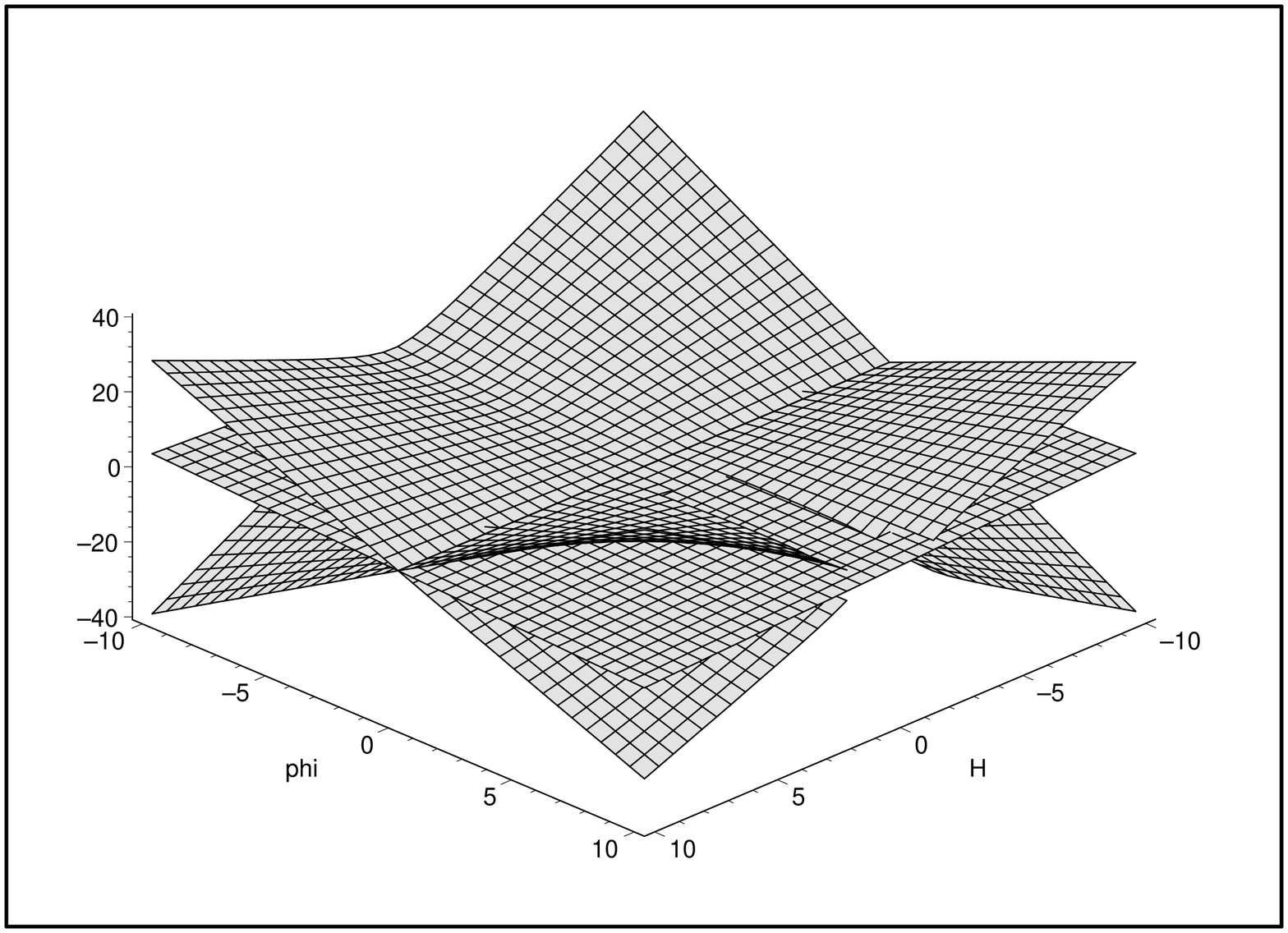}}}
\end{center}
\caption{The intersection between the ``energy surface'' and the ``ansatz 
surface''.}
\label{intersection}
\end{figure}

\section{Solutions of metric $f(R)$ gravity}
\label{sec:4}

$f(R)$ theories of gravity \cite{CCT, CDTT} are a 
subclass of scalar-tensor gravity with action
\be
S = \int d^4 x \, \frac{\sqrt{-g}}{16\pi}  \, f(R)  +S_{(m)} \,,
\label{f(R)action}
\ee
where $f({\cal R})$ is a non-linear function of the Ricci 
scalar $R$. This action is equivalent to a Brans-Dicke one. By defining 
the scalar field  $\phi = f'(R)$, it can be shown that the 
action~(\ref{f(R)action}) is equivalent to \cite{reviews1, reviews2, 
reviews3, CapozzielloFaraoni}
\be
S = \int d^4 x \, \frac{ \sqrt{-g}}{16\pi}  \left[ \phi R-V(\phi) \right]  
+S_{(m)} \,,
\ee
where
\be \label{f(R)potential}
V(\phi)= \phi R(\phi) -f\left(R(\phi)\right) \,,
\ee
and where  $ R=R(\phi) $ is now a function of $\phi=f'(R)$ usually
defined implicitly \cite{reviews1, reviews2, reviews3}. 
This theory has Brans-Dicke coupling $\omega=0$ and the 
potential~(\ref{f(R)potential}) for the Brans-Dicke 
scalar.

The Jordan frame solutions of Brans-Dicke cosmology reported in the 
previous sections can be seen also as solutions of some $f(R)$ cosmology.  
This is true if $\omega=0$ and 
\be
V_0 \left[ f'(R)\right]^{\beta}= R f'(R) -f(R) \,.
\ee
The functional form $f(R)=\mu R^n $, where $\mu$ and $n$ are constants,  
satisfies these requirements provided that\footnote{There is also a 
correspondence between solutions of $f(R)=R^n$ gravity and 
Einstein-conformally invariant Maxwell theory in $D$ dimensions 
\cite{Hendi}.} 
\be 
\beta=\frac{n}{n-1} \,, \label{beta}
\ee
\be
V_0=\frac{n-1}{n^{ \frac{n}{n-1}  } } \, \frac{1}{ \mu^{ \frac{1}{n-1} }}  
\,,\label{Vquesta}
\ee
for $n\neq 1$ (if  $n=1 $ this $f(R)$  theory reduces to GR). It must be 
$n>1$ to guarantee that $V_0>0$.  In practice, 
the value 
of $n$ is severely constrained by Solar System experiments, which require 
that $ n=1+\delta$ with 
$ \delta =\left( -1.1\pm 1.2 \right) \cdot 10^{-5} $   
\cite{SolarSystem1, SolarSystem2, SolarSystem3, 
SolarSystem4, SolarSystem5, newrefs1, newrefs2, newrefs3, 
newrefs4, newrefs5}. At 
the same time, any $f(R)$ 
theory must satisfy $f'>0$ in order for the graviton to 
carry positive energy and $f''\geq 0$ to guarantee local stability 
\cite{reviews1, reviews2,reviews3, mystabpaper}. These 
constraints are 
satisfied if $n=1+\delta $ with $\delta \geq 0$.   In spite of the 
experimental bounds on the exponent $n$, $R^n$ gravity has been the focus 
of much work aiming at exploring the possible phenomenology of $f(R)$ 
gravity and many phase space analyses for $R^n$ cosmology are available 
in the literature \cite{Rn1, Rn2, Rn3, Rn4, Rn5, Rn6, Rn7, 
Rn8, Rn9, Rn10, Rn11, Rn12, Rn13, Rn14, Rn15, Rn16, Rn17, 
Rn18, Rn19} (see \cite{Jose} for a phase space picture 
of general $f(R)$ cosmology analogous to that of the previous section). 
Moreover, in strong curvature regimes in the 
early universe, in which the present-day Solar System constraints do not 
apply, 
Starobinsky-like inflation \cite{Starobinsky} corresponding to $f(R)=R+\mu 
R^2$ 
is well 
approximated by $f(R) \simeq \mu R^2$.

When the conditions~(\ref{beta}) and 
(\ref{Vquesta}) 
are satisfied, the FLRW solutions of Brans-Dicke gravity with 
power-law potential reported in the previous sections are also solutions 
of $R^n$ gravity with or without a perfect fluid, which are added to the 
relatively scarce catalogue of exact 
solutions of these theories.

\section{Conclusions}
\label{sec:5}

The simple ansatz $\phi=\phi_0 a^p$ recovers most of the known solutions 
of Brans-Dicke cosmology and generates new ones in the presence of a 
power-law or inverse power-law potential, which is well-motivated in 
cosmology and particle physics. These solutions include power-law and 
exponential dependence of the scale factor $a(t)$ and of the 
Brans-Dicke field $\phi(t)$ from the comoving time $t$. This 
ansatz has a fairly  simple geometric interpretation  in the phase 
space 
of the solutions as the simultaneous search 
for a curve generated by the intersection of two surfaces and for the 
number $p$. The geometry of the phase space, 
however, can be complicated for various choices of the (inverse) 
power-law potential $V(\phi)=V_0\phi^{\beta}$ and if the 
spatial 
sections of the Brans-Dicke cosmology are curved. Details such as the 
integration constants appearing in the classic solutions as functions of 
the parameters of the theory have been provided by new general formulas, which 
are missing in the literature probably because of the non-availability of 
computer algebra at the time when these solutions were discovered. We have now a more 
unified and comprehensive view of analytical solutions of Brans-Dicke  
cosmology.

Prompted by the huge literature on $f(R)$ cosmology as an alternative to 
dark energy, it is natural to try to relate the old and new solutions of 
Brans-Dicke cosmology to corresponding solutions of $f(R)$ cosmology. It 
turns out that this is possible, and indeed relatively straightforward, 
for the family described by the choice $f(R)=\mu R^n$.  
The search for new $f(R)$ cosmologies will be continued in the future.

\section*{Acknowledgments}

We thank a referee for a useful discussion.  
D.K.C. thanks the Scientific and Technological Research Council of Turkey 
(T\"{U}B\.{I}TAK) for a postdoctoral fellowship through the Programme 
B\.{I}DEB-2219 and Nam{\i}k Kemal University for support.
V.F. is supported by the Natural Sciences and Engineering Research 
Council of Canada (2016-03803), and both authors thank Bishop's 
University.

\appendix\section{}
\label{appendix}

We have three different ways to balance the four terms in the Friedmann 
equation 
\be
\underbrace{ \frac{q^2}{t^2} \left( 1+p-\frac{\omega 
p^2}{6}\right) }_{(1)} 
+\underbrace{ \frac{k}{a_0^2\ t^{2q} }}_{(2)}
=\underbrace{ \frac{8\pi \rho_0}{3a_0^{p+3\gamma} \phi_0}  
\, \frac{1}{t^{q(p+3\gamma)}} }_{(3)} +
\underbrace{ \frac{V_0\phi_0^{\beta-1}}{6\ 
a_0^{p(1-\beta)}} \, 
\frac{1}{t^{pq(1-\beta)} } }_{(4)} \,: \label{FE} 
\ee
\begin{itemize}

\item $(1)$ balances $(3)$,  which gives $q(p+3\gamma)=2$, while $(2)$   
balances $(4)$, which yields $pq(\beta-1)=2q$. Therefore, it is  
\be
p=\frac{2}{1-\beta}\quad \mbox{and}\quad 
q=\frac{2(1-\beta)}{2+3\gamma(1-\beta)} \,,
\ee
and substituting these values into Eq.~(\ref{FE}), one obtains
\be
\Bigg\{ 
\frac{-2\omega+3(\beta-1)(\beta-3)}{\Big[2-3\gamma(\beta-1)\Big]^2}-\frac{2\pi\rho_0\,a_0^{\frac{2-3\gamma(\beta-1)}{\beta-1}}}{\phi_0 
}\Bigg\}\frac{4}{3t^2} 
=\frac{\Big(V_0\,\phi_0^{\beta-1}-6k\Big)}{6a_0^2} 
\frac{1}{t^{\frac{-4(\beta-1)}{2-3\gamma(\beta-1)}}} \,. 
\ee
Equating to zero the terms in parenthesis separately yields
\begin{eqnarray}	
\rho_0&=&\frac{-2\omega+3(\beta-1)(\beta-3)}{2\pi\Bigg[ 
2-3\gamma(\beta-1)\Bigg]^2}\phi_0 \, 
a_0^{\frac{-2+3\gamma(\beta-1)}{\beta-1}} \,,\\
&&\\ \nonumber
k&=&\frac{V_0\phi^{\beta-1}}{6} \,.
\end{eqnarray}
If we substitute these $\rho_0$ and $k$ values into the 
acceleration equation (\ref{6}), we obtain
\be
\left( -3+2\omega+3\beta \right) 
\Bigg\{\frac{12\big[2-4\beta+2\omega(\gamma-2) 
+3\gamma(\beta-1)\big]}{(2-3\gamma(\beta-1))^2t^2} 
+\frac{V_0\phi_0^{\beta-1}}{a_0^2t^{\frac{-4(\beta-1)}{2-3\gamma(\beta-1)}}} 
\Bigg\}=0 \,.
\ee
This equation can be satisfied in two ways: the first one consists of 
setting the first parenthesis to zero, while the second way consists of  
setting the second parenthesis to 
zero. However, the second possibility is 
already discussed in Sec.~\ref{subsec:2.7}. Setting the first parenthesis 
to zero gives
\be
\beta=\frac{3-2\omega}{3} \,.
\ee
Now we need to satisfy the scalar field equation~(\ref{7}),  
which gives
\be
\frac{2a_0^2(2\omega+3)}{t^2}-\frac{V_0\, 
\phi_0^{-\frac{2\omega}{3}}(\omega\gamma+1)^2}{ 
t^{\frac{4\omega}{3(\omega\gamma+1)}}}=0 \,,
\ee
requiring one to set the powers of $t$ equal to each other. Therefore, we  
conclude that balancing these two pairs does not produce any new solution.
	
\item Balancing the terms $(2)$ and $(3)$ gives $2q=q(p+3\gamma)$, whereas 
balancing $(1)$ and  $(4)$ yields $pq(1-\beta)=2$, therefore
\be
p=2-3\gamma \quad \mbox{and}\quad  q=\frac{2}{(3\gamma-2)(\beta-1)} \,,
\ee
while the Friedmann equation~(\ref{FE}) gives
\be
\Bigg\{\frac{4[\omega(3\gamma-2)^2+18(\gamma-1)]}{ 
(3\gamma-2)^2(\beta-1)^2}+\frac{V_0\, 
\phi_0^{\beta-1}}{a_0^{(\beta-1)(3\gamma-2)}} \Bigg\} 
\frac{1}{t^2}=\frac{3k\,  \phi_0-8\pi\, 
\rho_0}{a_0^2\phi_0}\frac{2}{t^{\frac{4}{(\beta-1)(3\gamma-2)}}} \, ,
\ee
which implies that
\begin{eqnarray}
V_0&=&-\frac{4\, a_0^{(\beta-1)(3\gamma-2)}[ 
\omega(3\gamma-2)^2+18(\gamma-1)]}{\phi_0^{\beta-1}(3\gamma-2)^2 
(\beta-1)^2} \,,\\
&&\nonumber\\
k & = & \frac{8\pi\,\rho_0}{3\phi_0} \,.
\end{eqnarray}
If we substitute these values of $V_0$ and $k$ into the 
acceleration equation~(\ref{6}), we obtain
\be
\left[ 3+\omega(3\gamma-2) \right]\Bigg\{\frac{6(2\omega+2\beta-1) 
-9\gamma(2\omega+\beta+1)}{(3\gamma-2)^2(\beta-1)^2\, 
t^2}-\frac{4\pi\, \rho_0}{a_0^2\phi_0\, 
t^{\frac{4}{(\beta-1)(3\gamma-2)}}}\Bigg\}=0 \,.
\ee
A possible solution would be obtained if the prefactor 
$ \left[ 3+\omega(3\gamma-2) \right]$ vanishes, giving
\be
\gamma=\frac{2\omega-3}{3\omega} \,.
\ee
In order to satisfy the scalar field equation~(\ref{7}), we substitute 
the values of $V_0$,  $k$, and $\gamma$ to obtain 
\be
\frac{\omega(2\omega+3)(\beta+1)}{t^2}-\frac{12\pi\, 
\rho_0(\beta-1)^2}{a_0^2\, \phi_0\,t^{\frac{-4\omega}{3(\beta-1)}}}=0 \,.
\ee
Finding a solution without setting the powers of $t$ equal to each 
other requires, at a minimum, to set $\beta=1$, which makes 
the power of $t$ infinite. Therefore, this choice of balancing terms give 
no reasonable solution without setting the powers of $t$ equal to 
each other.
	
\item  Another possible solution of the dynamical equations arises if 
$(1)$ balances $(2)$, which gives $2q=2$, while  $(3)$ balances $(4)$, 
which yields  $q(p+3\gamma)=pq(1-\beta)$. Therefore $q$ and $p$ 
become
\be
q=1\quad \mbox{and}\quad p=-\frac{3\gamma}{\beta} \,.
\ee
Substituting these values of $q$ and $p$ into the Friedmann 
equation~(\ref{FE}) leads to
\be
\Bigg[\frac{6k}{a_0^2}-\frac{3(3\omega\gamma^2
+6\gamma\beta-2\beta^2)}{\beta^2 }\Bigg] \frac{1}{t^2}
=\Bigg[\frac{16\pi \,\rho_0}{\phi_0}+V_0\phi_0^{\beta-1}\Bigg] 
\frac{1}{(a_0\, t)^{\frac{3\gamma(\beta-1)}{\beta}}} \,,
\ee
which requires
\begin{eqnarray}
k&=&\frac{a_0^2\left( 3\omega\gamma^2+6\gamma\beta -2\beta^2 
\right)}{2\beta^2} \,,\\
&&\nonumber\\
\rho_0 &=& -\frac{V_0\phi_0^\beta}{16\pi} \,.
\end{eqnarray}
Requiring that the energy density and the potential energy  
density be non-negative, one must set 
$\rho_0=V_0=0$. Substituting these values of $k$, $\rho_0$, and $V_0$ 
into the acceleration equation (\ref{6}) leads to 
\be
3\gamma(\omega\gamma+\beta)=0 \,,
\ee
and $\gamma$ becomes 
\be
\gamma=0\quad \mbox{or}\quad \gamma=-\frac{\beta}{\omega} \,.
\ee
We still have  to satisfy the scalar field equation~(\ref{7}). If we 
set  $\gamma=0$, the scalar field becomes constant, $\phi(t)=\phi_0$, 
which reduces the context to GR. For $\gamma=- \beta/\omega$ one 
obtains instead 
\be
\frac{3a_0^{3/\omega} \left( 2\omega+3 \right)\phi_0}{ 
t^{\frac{2\omega-3}{\omega}} \, \omega^2}=0 \,,
\ee
which is satisfied only if $\omega=-3/2$, the unacceptable  value of 
the Brans-Dicke parameter ruled out from the beginning.

\end{itemize}

These three cases show that making different matches of the terms in  
Friedmann equation~(\ref{FE}) does not yield new solutions. 

\section*{Acknowledgments}

D.K. thanks the Scientific and Technological Research Council of Turkey 
(T\"{U}B\.{I}TAK) for a postdoctoral fellowship through the Programme 
B\.{I}DEB-2219 and Nam{\i}k Kemal University for support.
V.F. is supported by the Natural Sciences and Engineering Research 
Council of Canada (2016-03803), and both authors thank Bishop's 
University.

\end{document}